\documentclass[useAMS,usenatbib,usegraphicx]{mn2e}

\usepackage{times}
\usepackage{subfigure}

\newcommand{\kms}{\hbox{km s$^{-1}$}}
\newcommand{\ms}{\hbox{m s$^{-1}$}}

\newcommand{\vsini}{\hbox{$v$\,sin\,$i$}}

\newcommand{\degs}{$\degr$}
\newcommand{\chisq}{$\chi^{2}$}

\title[Near infrared spectroscopic search for HD 75289b]{Near infrared spectroscopic search for the close orbiting \newline planet \hbox{HD 75289b}}
\author[J.R.~Barnes, C.J.~Leigh, H.R.A.~Jones, Travis~S.~Barman, D.J.~Pinfield, A.~Collier~Cameron, J.S.~Jenkins]
{J.R.~Barnes$^1$\thanks{E-mail: j.r.barnes@herts.ac.uk} 
 C.J.~Leigh$^2$, 
 H.R.A.~Jones$^1$, 
 Travis~S.~Barman$^3$, 
 D.J.~Pinfield$^1$,
 \newauthor
 A.~Collier~Cameron$^4$ and
 J.S.~Jenkins$^1$ \\
$^1$ Centre for Astrophysics Research, University of Hertfordshire, Hertfordshire AL10 9AB. UK. \\ 
$^2$ Astrophysics Research Institute, Liverpool John Moores University, Birkenhead CH41 1LD. UK. \\
$^3$ Lowell Observatory, Planetary Research Centre, 1400 West Mars Hill Road, Flagstaff, AZ 86001. USA \\
$^4$ SUPA, School of Physics and Astronomy, University of St Andrews, Fife KY16 9SS. UK. \\
}

\begin{document}

\date{Received 12/03/2007; Revised 02/05/2007.}


\maketitle

\label{firstpage}

\begin{abstract}

{We present a search for the near infrared spectroscopic signature of the close orbiting extrasolar giant planet \hbox{HD 75289b}. We obtained $\sim$230 spectra { in the wavelength range \hbox{$ 2.18$\ \micron\,-\,$ 2.19$\ $\micron$}} using the Phoenix spectrograph at Gemini South. By considering the direct spectrum, derived from irradiated model atmospheres, we search for the absorption profile signature present in the combined star and planet light. Since the planetary spectrum is separated from the stellar spectrum at most phases, we apply a phase dependent orbital model and tomographic techniques to search for absorption signatures.

Because the absorption signature lies buried in the noise of a single exposure we apply a multiline deconvolution to the spectral lines available in order to boost the effective S/N ratio of the data. The wavelength coverage of 80~\AA\ is expected to contain $\sim100$ planetary lines, enabling a mean line with S/N ratio of 800 to be achieved after deconvolution. We are nevertheless unable to detect the presence of the planet in the data and carry out further simulations to show that broader wavelength coverage should enable a planet like \hbox{HD 75289b} to be detected with 99.9 per cent (4 $\sigma$) confidence. We investigate the sensitivity of our method and estimate detection tolerances for mismatches between observed and model planetary atmospheres.}

\end{abstract}


\section{INTRODUCTION}
\protect\label{section:intro}

One of the great surprises in the search for worlds orbiting other stars was the discovery of a class of close orbiting Extrasolar Giant Planets (CEGPs). The existence of such objects did not fit with the formation scenarios used to explain the arrangement of planets in our own solar system. Even before the first extrasolar gaseous giants were discovered, core accretion models predicted that they should only form \citep{pollack84review} at distances similar to those of the giant outer planets \citep{boss95jupiters}. Here temperatures must be cool enough to enable the formation of an ice (water, methane and ammonia in solid or liquid form) core from the proto-planetary disk before gas accretion onto the core can then take place.

With the announcement of the discovery of the first CEGP orbiting the F8 dwarf 51 Peg \citep{mayor95} at a distance of only 0.05 AU came the realisation that existing theories of planetary formation needed to be revised. \citet{guillot96} argued, contrary to conventional belief \citep{boss95jupiters}, that gas giant planets could survive in such close proximity, although their formation beyond the ice point was still required. Mechanisms to enable an exchange of angular momentum to take place in a proto-planetary disk are believed to be responsible for the spiralling into and eventual halting of planets in a close orbit \citep{lin96}.

Planetary atmospheric physics research in recent years has been motivated by the expectation that radiation processed by a large body close to its parent star should be detectable with modern instrumentation. The first clear detection of sodium in the atmosphere of \hbox{HD 209458b} \citep{charbonneau02atmos} revealed a lower than expected concentration when compared with a cloudless planetary atmosphere. Several explanations for the lack of sodium were put forward, including the possible presence of a high cloud deck. No observations had been able to differentiate between the possibility of no clouds (combined with lower sodium abundance) until recent Spitzer observations of HD 209458b with the low resolution spectrometer IRS \citep{houck04irs} indicated a spectral feature consistent with a high silicate cloud deck \citep{richardson07spectrum}.

{ The reflected light spectroscopic studies carried out by \citep{cameron99tauboo,charbonneau99tauboo,cameron02upsand,leigh03a,leigh03b} and more recent results from MOST photometry \citep{rowe06} place albedo {\em upper limits} of 0.1\,-\,0.25 on the atmospheres of CEGPs. \citet{leigh03b} has placed an upper limit of 0.12 on the geometric albedo of \hbox{HD 75289b}, suggesting that if clouds are present at all, they are highly non-reflective. These are somewhat lower than the solar system gas giants, Jupiter, Saturn, Uranus and Neptune, which possess geometric albedos of 0.46, 0.39, 0.60 and 0.58 respectively \citep{karkoschka94}. These observed albedo limits for CEGPs rule out the high silicate cloud investigated in the models of \citet{sudarsky00}, and contradict more recent Spitzer observations \citep{richardson07spectrum}. }


A number of attempts to model the emergent spectrum of CEGPs have also been made in recent years. \citet{sudarsky03} have calculated spectra for planets with a range of orbital separations from the parent star, including the so called class of `hot roasters'. These authors also presented spectra in the 0.4\,-\,5 \micron\ region for specific systems. For the well studied transiting planet, \hbox{HD 209458b}, they found a phase averaged planet-star flux ratio of log$_{10}(F_p/F_*)\sim -3.10$. A re-evaluation of the \citet{sudarsky03} models by \citet{richardson03} which include phase dependent effects have yielded 2.2 \micron\ flux ratios of up to log$_{10}(F_p/F_*)\sim -2.89$ for a cloudless atmosphere in the case of \hbox{HD 209458b}. Here, the incident radiation is completely absorbed and re-emitted on the day-side of the planet. A more recent study by \citet{bha05} (BHA05), which models day-night gradients, has yielded results consistent with previous studies. For no redistribution of heat (i.e. re-emission of radiation on the day side), they find log$_{10}(F_p/F_*)\sim -2.90$, while uniform redistribution of the incident radiation yields log$_{10}(F_p/F_*)\sim -3.05$ at all phases for \hbox{HD 209458b}.

Several pioneering attempts were made to detect CEGPs in the near infrared (NIR) from the ground in the combined star and planet light \citep{wiedemann01,lucas02,snellen05}. Success in the IR was achieved with the Spitzer Space Telescope, with detections of a reduction in thermal emission during secondary transits of \hbox{HD 209458b} { (G0V)} \citep{deming05hd209458} and TrES-1 { (K0V)} \citep{charbonneau05tres1}. \hbox{HD 209458b} photometry indicates log$_{10}(F_p/F_*)\sim -2.59$ at 24 \micron\ while a 2.3 \micron\ ground based measurement yields a value of log$_{10}(F_p/F_*)\sim -3.00$, albeit with a large degree of uncertainty \citep{snellen05,deming06ir}. The TrES-1 photometry \citep{charbonneau05tres1} indicated log$_{10}(F_p/F_*)\sim -3.15$ at 4.5 \micron\ and log$_{10}(F_p/F_*) \sim -2.66$ at 8.3 \micron. These observations are consistent with the \citet{sudarsky03} base model. { \citet{deming06hd189733b} found that the CEGP \hbox{HD 189733b} yielded an even higher contrast ratio (log$_{10}(F_p/F_*)\sim -2.26$ at 24 \micron) due to the close orbit and later spectral type of the K2V star. Fig. 8 of \citet{fortney06hd149026b} plots a comparison of observed and predicted star-planet contrast ratios covering the spectral region \hbox{3 \micron\,-\,30 \micron}}. The \hbox{HD 209458b} and TrES-1 observations are consistent with the models, with a 2.2 \micron\ signal strength of 0.1\%, whereas scaling the \hbox{HD 189733b} signal suggests a signal strength as high as 0.2\%. Most recently, \citet{harrington06} have used high S/N Spitzer photometric observations to directly measure the planetary flux of the CEGP $\upsilon$ And at 24 \micron. They find a phase dependent flux which is consistent with the BHA05 models where heat is not significantly redistributed throughout the atmosphere but re-radiated on the dayside of the planet \hbox{HD 209458b}.

\subsection{\hbox{HD 75289}}
\protect\label{section:hd75289}

A companion to the main sequence G0 dwarf \hbox{HD 75289} was first announced by \citet{udry00} following precision radial velocity monitoring with the CORALIE spectrometer at the 1.2-m Euler Swiss telescope. Since its first discovery, \citet{pepe02} have published a refinement of the system parameters in light of further observations. A comprehensive list of system parameters have been estimated and tabulated by \citet{leigh03b}. Subsequent to further refinement of parameters, as a result of a longer timebase of observations, \citet{butler06catalogue} give an ephemeris of \hbox{$\phi=2450829.872 + 3.509267E$ day}. This result, with \hbox{$P = 3.509267 \pm 0.000064$~day}, is marginally greater than the \hbox{$P = 3.5091 \pm 0.0001$~day} estimate of \citet{udry00}. \citet{butler06catalogue} estimate an orbital radius semi-major axis of \hbox{$a = 0.0482$ AU} while \citet{udry00} estimate \hbox{$a =0.0483$ AU}. 

\begin{figure}
\begin{center}
\includegraphics[height=8.75cm,angle=270]{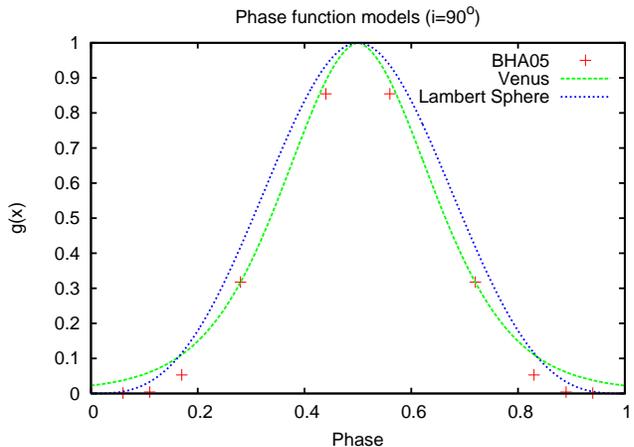}
\caption{Phase function, $g(\alpha,\lambda)$, representing three different models with an inclination of $i=90$\degs. Plotted are an empirical reflected light Venus \citep{hilton92} phase function, a Lambert Sphere phase function and the 2.2 \micron\ phase function derived from the atmospheric models of BHA05. In the phase range \mbox{$0.28 \leq \phi \leq 0.72$}\,, the BHA05 model closely mimics the Venus phase function model.}
\protect\label{fig:phasfunc}
\end{center}
\end{figure}

\begin{table*}
\caption{Journal of observations for 2006 January 15 \& 16 Gemini South - Phoenix observations. Observations were made in an ABBA sequences. Each individual exposure was created from coadding shorter exposures. For HD 75063, the mean S/N ratio for each co-added group of spectra is given in column 5. The combined HD 75063 spectra were used to remove the telluric features from each AB or BA combined pair of HD 75289 spectra post extraction (see \S \ref{section:extracting}). The mean S/N ratio after telluric lines and stellar lines were removed is given in column 5 for HD 75289.}
\protect\label{tab:journal}
\vspace{5mm}
\begin{center}
\begin{tabular}{lccccc}
\hline
Object			& UT start	& Exp time per	& Number of ABBA	& Mean	& Comments  \\
			& 		& exposure [s]	& sequences		& S/N	& 	    \\
\hline
\multicolumn{5}{c}{2006 January 15} \\
\hline
HD 75289		& 01:46:55	&  120		&	6		& 401	& Seeing = 0.68\arcsec \\
HD 75063		& 02:50:32	&   40   	&	2		& 391	& G5V Telluric standard \\
HD 75289		& 03:18:48	&  120 		&	12		& 133	& Slit mis-alignment (counts decreasing)  \\
HD 75063		& 05:22:47	&   40   	&	3		& 398	&  \\
HD 75289		& 06:03:57	&  120 		&	12		& 380	&  Low counts: slit re-alignment in middle of set. \\
HD 75063		& 08:46:42	&   40   	&	3		& 515	&  \\
\hline
\multicolumn{5}{c}{2006 January 16} \\
\hline
HD 75063		& 00:48:25	&   40   	&	3		& 414	&  \\
HD 75289		& 01:14:02	&  160 		&	10		& 358	&  \\
HD 75063		& 03:32:38	&   40   	&	3		& 240	&  \\
HD 75289		& 03:59:03	&  160		&	11		& 363	& Software crash at end of sequence: frames rejected \\
HD 75063		& 06:25:46	&   40   	&	3		& 385	&  \\
HD 75289		& 03:48:51	&  160		&	11		& 337		&  \\
HD 75063		& 09:06:56	&   40   	&	3		& 405	&  \\

\hline
\end{tabular}
\end{center}
\end{table*}

We have carried out Monte-Carlo simulations to determine the most probable estimates of planetary mass, $M_p$, and orbital inclination, $i$. { An estimate of the orbital inclination provides an estimate for the most probable velocity amplitude of the planetary signal. As described in \S \ref{section:results} we can use this value as a guide when searching for a planetary signal.} The axial inclination of the system can simply be calculated from the equatorial rotation velocity, $v_{eq}$, determined from estimates of the stellar radius and rotation period, while \vsini~can be measured directly. We assume that the stellar equator and planetary orbit lie in the same plane in agreement with Rossiter-McLaughlin effect measurements for the HD 209458 \citep{bundy00hd209458, queloz00hd209458, winn05, wittenmyer05} and HD 189733 \citep{winn06} systems. Monte-Carlo simulations using updated values discussed above and values tabulated in Table 1 of \citet{leigh03b} yield a most probable value of $i = 67$\degs. A planetary mass of \hbox{$M_p = 0.51 \pm 0.08 M_J$} then follows from the known orbital amplitude of the stellar reflex motion of \hbox{$K_* = 54.9 \pm 1.8~$\ms.}

In light of observational and theoretical results, the adopted theoretical $1.6 R_J$ radius determined by \citep{leigh03c} is probably a significant over-estimate. A recent plot \citep{burrows06radii} of radii determined from all observations of transiting CEGPs indicates \hbox{$R \sim 1\,-\,1.4 R_J$}. 
Based on the recent theoretical evolution models from Fig. 3 of \citet{burrows06radii} for CEGP radii, we estimate an approximate radius of $R_p \sim 1.2~R_J$.

In this work, we aim to detect the planetary signature of the close orbiting planet \hbox{HD 75289b}. In the NIR, rather than searching for an attenuated copy of the stellar spectrum, we can search for the direct spectrum emerging from the heated atmosphere which is expected to possess temperatures in the range 1000-1200 K \citep{deming05hd209458,charbonneau05tres1,deming06hd189733b}. { The analysis presented below relies on the use of the predicted high density of absorption lines, due mainly to H$_2$O, to search for the faint planetary signal. As such it has the potential to provide constraints on the reliability of model atmosphere opacities.} In \S \ref{section:detect} we present the method used to detect a NIR planetary signal while \S \ref{section:obsdata} outlines the data reduction and analysis of data. We discuss our results in \S \ref{section:results} and carry out further simulations in \S \ref{section:simplanet} before discussing future prospects for this kind of survey in \S \ref{section:discussion}.

\section{Detecting a Near Infrared spectroscopic signature}
\protect\label{section:detect}

\subsection{Phase Function}

The method of analysis uses a modification of the matched filter technique first presented in \citet{cameron99tauboo}. This technique has been refined and applied to several systems using optical data \citep{cameron02upsand,leigh03a,leigh03c} including \hbox{HD 75289b} \citep{leigh03b} where an upper limit of log$_{10}(F_p/F_*)=-4.52$ has been determined. Here we seek to apply a similar method to NIR spectra.

In order to extract the planetary signal from a timeseries of spectra, we model the planetary signature as a phase dependent spectrum superimposed on an unvarying stellar spectrum. The time dependent variations of the planet orbiting the star are: (1) the instantaneous Doppler shift of the spectrum due to the relative position of the planet and (2) a phase dependent planet-star flux ratio, $F_p/F_*$, which is dependent on atmospheric physics and heating due to the parent star.

The ephemeris and velocity semi-amplitude, \mbox{$K_{p} = v_{p} sin i =$}~\hbox{$137.6$~\kms} (where $i$ is the orbital inclination), enable the instantaneous velocity shift of the planetary spectrum relative to the stellar spectrum to be determined at any observation phase. Since the rotation velocity of \hbox{HD 75289} is only \hbox{3.8 \kms}~the planetary signature will be Doppler shifted clear of the stellar lines at all phases except those close to $\phi = 0.0~\&~0.5$. A clean spectral separation will thus be present at most phases.

Because we observe the combined light from star and planet, the light is dominated by the former, meaning that we must express the planetary signature as a fraction of the stellar signature. The planet-star flux ratio, $\epsilon$ can be expressed as a function of orbital phase ($\phi$) and wavelength ($\lambda$), such that

\begin{equation}
\epsilon(\alpha,\lambda) \equiv \frac{f_{p}(\alpha,\lambda)}{f_{*}(\lambda)} = \epsilon_{0}(\lambda) g(\alpha,\lambda).
\end{equation}

The form of this function is similar to that used by in optical studies (e.g. \citep{cameron02upsand}) but with the geometric terms describing the albedo, planetary radius and orbital semi-major axis combined into a single function $\epsilon_0(\lambda)$. The value of $\epsilon_{0}(\lambda)$ can be derived from model spectra at the desired value of $\lambda$ and represents the {\em maximum} planet-star flux ratio, observed at phase 0.5, when $i = 90$\degs. The phase angle $\alpha$ combines the orbital inclination, $i$, and orbital phase effects and is defined as

\begin{equation}
cos(\alpha) = -sin(i) cos(\phi).
\end{equation}

The phase function $g(\alpha,\lambda)$ may also either be empirically determined or modelled.

{ For a cloud free model, \citet{marley99} showed that planets reflect most efficiently shortward of \hbox{$\sim6000$ \AA}, where photons undergo Rayleigh scattering before being absorbed. \citet{seager00} found that the form of the phase function is strongly dependent on the particle size at 5500 \AA, with larger particles giving strong back scattering and albedos peaking at values similar to the Jovian planets in out own solar system.}

We have investigated the form of the phase function for \hbox{HD 209458} based on the results of BHA05. Fig. \ref{fig:phasfunc} is a plot of $g(\alpha,\lambda)$ for the case where $i = 90$\degs\ showing the close similarity of the Venus phase function and BHA05 phase functions for \hbox{$\phi$ = 0.25\,-\,0.75.} Since we are concerned with the detection of the planetary signal and not the characterisation (i.e. we are no attempting to distinguish between phase function forms) in this paper, we have adopted the Venus phase function in subsequent analyses. This function has the advantage of being able to mimic inclination effects through use of the $\alpha$ parameter, not possible with our empirical function, and is a valid approximation as our spectra were obtained with phases \mbox{$0.275 \leq \phi \leq 0.72$}.

\section{Observations \& Data Reduction}
\protect\label{section:obsdata}

We present the observations made with the Phoenix spectrograph { \citep{hinkle03phoenix}} at Gemini South on 2006 January 15 \& 16. Densely sampled timeseries were recorded using the 256 $\times$ 1024 InSb Aladdin II array. In addition to the \hbox{HD 75289} spectra, observations of the bright A1 III star, \hbox{HD 75063}, were made to enable careful monitoring of atmospheric absorption features. Because of vignetting and cosmetic issues, the useful region of spectrum was trimmed in the dispersion direction during processing to give a useful area of \mbox{$256 \times 850$}. The spectral range of 81.74 \AA\ covers the 21807.32 \AA\ to 21889.06 \AA\ (2.18 \micron\,-\,2.19 \micron) region of the NIR at a spectral resolution of \hbox{56\,800}. With the 1024 pixel array, this gives a mean pixel resolution of 0.096 \AA\ per pixel which is equivalent to a 1.32 \kms~velocity increment at the central wavelength. The journal of observations is recorded in Table \ref{tab:journal}.

\subsection{Detector}
\protect\label{section:detector}
The Aladdin array suffers from a number of significant cosmetic defects. These include a number of dust spots and elongated features. Hot pixels are also present and were flagged during the extraction process to ensure that they were not included. The left side of the detector appears to exhibit sensitivity which alternates between adjacent rows. This stripe pattern produces a ripple at the $\sim$4\% level but is not present on the right hand portion of the detector.

The characteristics of the Aladdin detector also necessitated an observing strategy to monitor the faint residual charge persistence signature common in this kind of detector when working in the infrared \citep{solomon99thesis}. Observations were thus made by alternating the position of the star on the slit and thus the spectrum on the detector in an ABBA type pattern to allow any persistence signature to be monitored. Each of the 4 sub-exposures in an ABBA sequence comprised a number of further sub-exposures. Short exposure times help to { minimise dark} current and may help to reduce persistence levels. For 40s exposures, 2x20s exposures were used. Similarly, 120s = 2x60s and 160s = 4x40s. \citet{randall06} has however found that continual flushing of the array can increase the dark current floor, and {\em increase} the persistence current. 

\begin{figure}
\begin{center}
\vspace{1.0cm}
  \includegraphics[height=7.75cm,angle=270,bbllx=121,bblly=94,bburx=490,bbury=699]{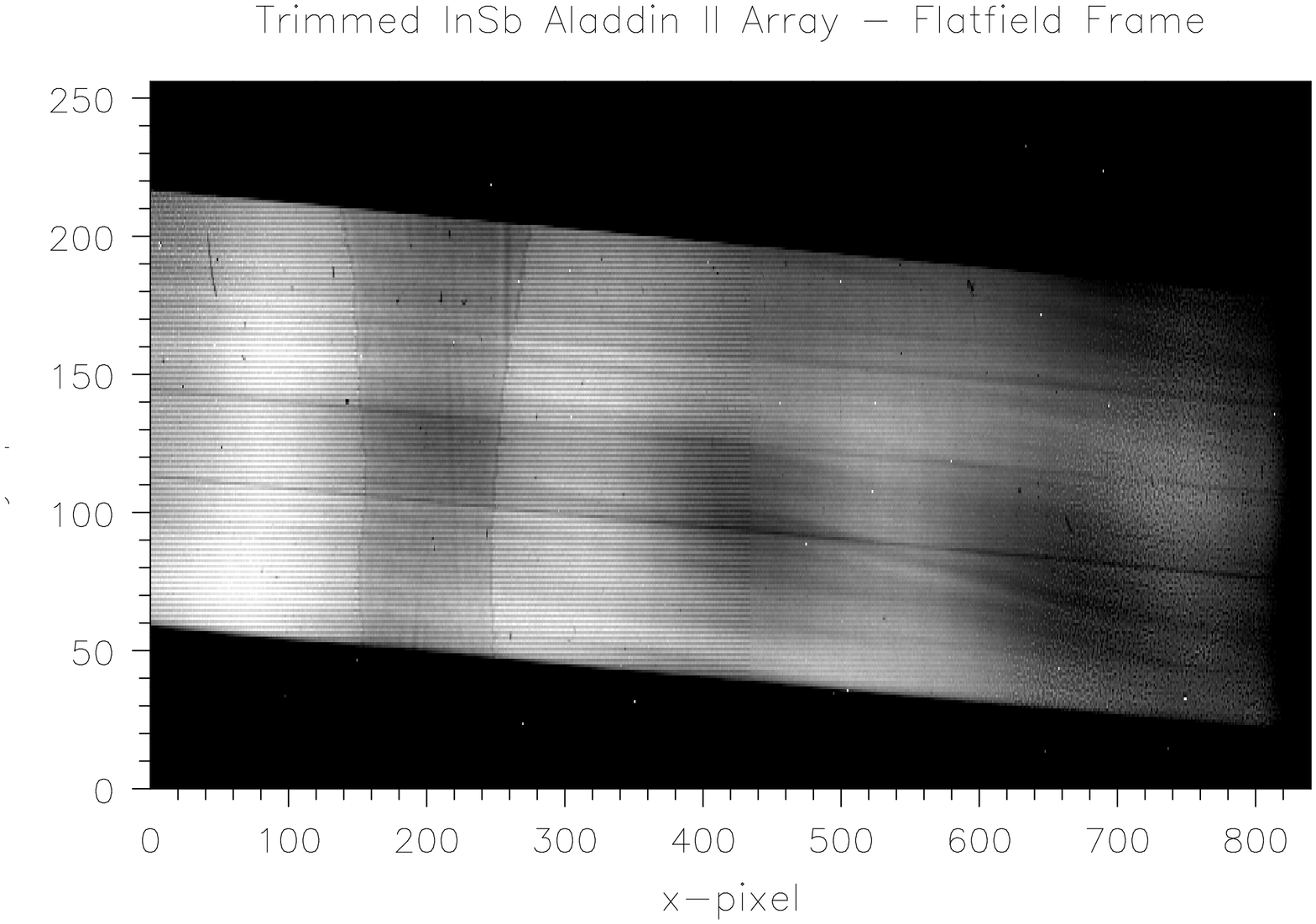}
\vspace{1.5cm} \\
  \includegraphics[width=3.25cm,height=7.75cm,angle=270,bbllx=81,bblly=95,bburx=530,bbury=699]{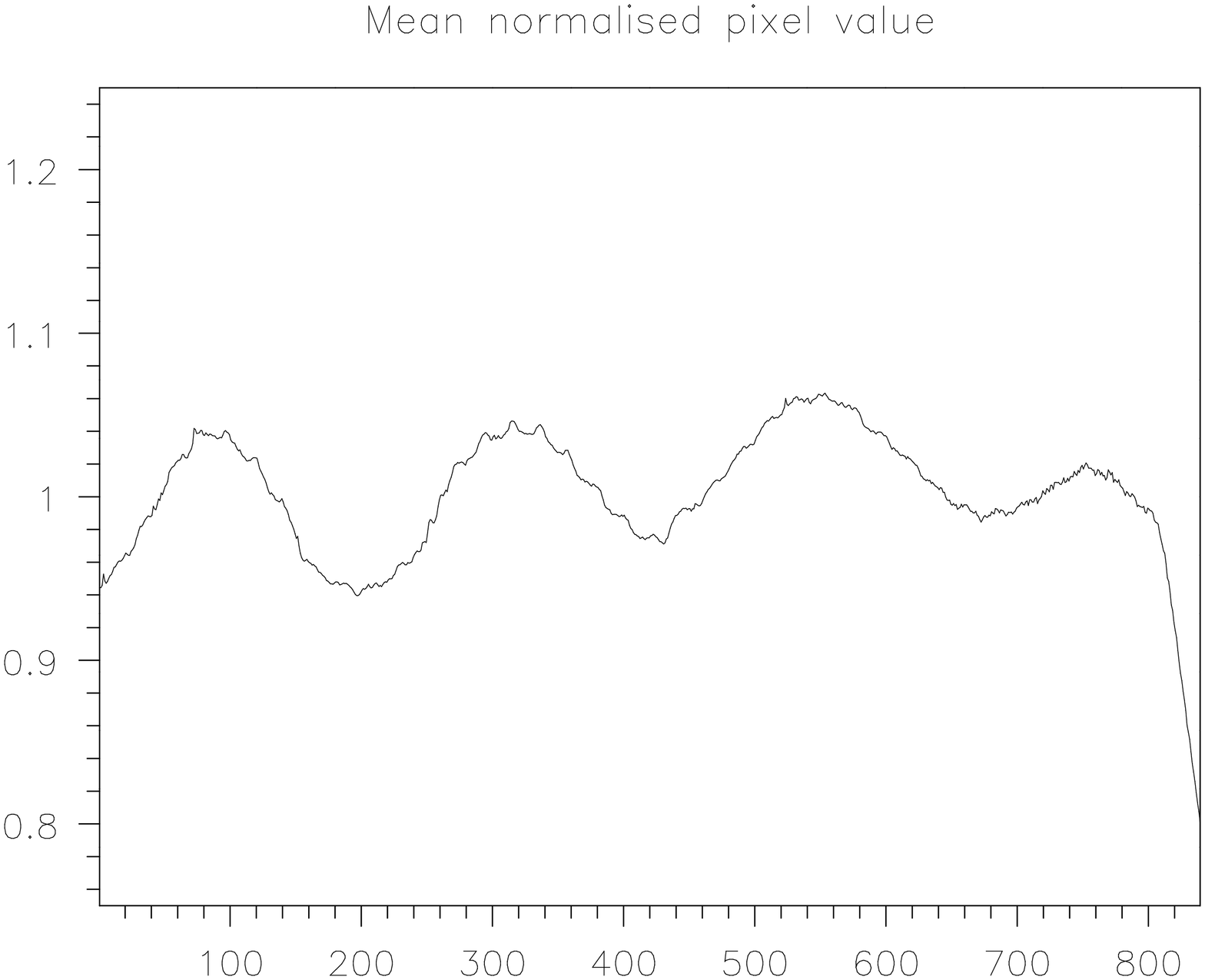}
\vspace{1cm} \\
\end{center}

\caption{Top: Aladdin II array master flatfield image showing bad and hot pixels, the large scale variations and the row to row pixel sensitivity alternation on the left had part of the detector. The large scale variations were removed post-extraction by fitting splines to the continuum. The greyscale runs from black=0.75 to white=1.25. Bottom: The mean profile of the largescale variations created after rotating the frame (to ensure the spectrum ran parallel to the detector) and collapsing the spectrum. The spectrum was then normalised by dividing by a straight line which was used to remove the variation in throughput of the spectrograph (from left to right). The plot shows that the mean variation of the large scale ripples is expected to be at the 5\% level in the extracted spectra.}
\protect\label{fig:ccd_flatfield}
\end{figure}

We could not detect persistence of the \hbox{HD 75289} spectra which typically peaked at { 5000 e$^-$s}. The \hbox{HD 75063} spectra however yielded a maximum of { 12500 e$^-$s} and did leave a faint persistence trace at a level of 1.6$\sigma$ of the sky background or 0.13\% of the peak counts when switching between A and B positions. For \hbox{HD 75063}, telluric lines may shift wavelength throughout the night by a few tens of \ms~\citep{gray06tellurics}, but this effect translates to a shift of 1/130th of a pixel. Hence the persistence effect when coadding these spectra is negligible. The effect is more crucial, if significant variation in the position of features is expected, such as from an orbiting planet whose signature is Doppler shifted during its orbit. If a signal of similar relative strength to that in the \hbox{HD 75063} spectra is seen in the \hbox{HD 75289} spectra, the blurring effect is still expected to be a second order phenomenon and we do not expect to detect such a signature.

\subsection{Data Reduction}
\protect\label{section:data_reduction}
Pixel to pixel variations were removed using flat-field exposures taken with an internal tungsten reference lamp. In order to create a reliable balance frame to remove the pixel sensitivity variations, we divided a Gaussian blurred (using a FWHM of 7 pixels) version of the master flat field image by the original master flat field image. Fig. \ref{fig:ccd_flatfield} (top) shows the flatfield image normalised to unity. The large-scale variations at the 5 per cent level shown in Fig. \ref{fig:ccd_flatfield} (bottom) are a consequence of broad sensitivity variations on the Aladdin II array. We found that these variations were dependent on the illumination of the array, { being both source and time dependent,} resulting in an inability to remove the pattern using standard flatfielding techniques. For example, we extracted the flat-field variation with a profile used for extraction of our object frames. The morphology of the flatfield ripples was found to differ from the continuum ripple in the extracted object frame. Similarly, since the ripple pattern was source dependent, we were unable to flatfield the spectra by dividing by the standard star, \hbox{HD 75063}. { While the pattern for a given object remained stationary to first order throughout each night of observations, second order time dependent variation was also seen. Nodding the telescope (see below) between two positions on the slit resulted in a drift of the A and B positions of several pixels throughout the night. During extraction we therefore traced each spectrum independently. Since the ripple pattern described above appeared different for HD 75289b than for the flat field, we removed it by fitting splines to the continuum of the extracted spectra.} 

The worst cosmic ray events were removed at the pre-extraction stage using the Starlink {\sc figaro} routine {\sc bclean} \citep{shortridge93figaro}. { Instead of making use of the ABBA sequences to reject sky lines by extracting from the A-B and B-A differences \citep{joyce92irarrays}, we found that the S/N ratios of the extracted spectra were optimised when the sky background was modelled by fitting polynomials of degree 3 to the pixels either side of the spectral profile at each X-position in each frame. An iterative fitting was used to reject deviant X-position sky fits, thereby rejecting any sky lines. Only a single sky line at $\sim21868$ \AA\ was present in the data at a level of $\sim9$ per cent in the worst spectra (on January 15) and 1 per cent in a typical spectrum.}. The spectra were extracted using {\sc echomop}, the \'{e}chelle reduction package developed by \citet{mills92}. The spectra were extracted using {\sc echomop}'s implementation of the {\sc optimal} extraction algorithm developed by \citet{horne86extopt}. {\sc echomop} propagates error information based on photon statistics and readout noise throughout the extraction process. \\

\begin{figure}
\begin{center}
\vspace{1.0cm}
\includegraphics[height=85mm,angle=270]{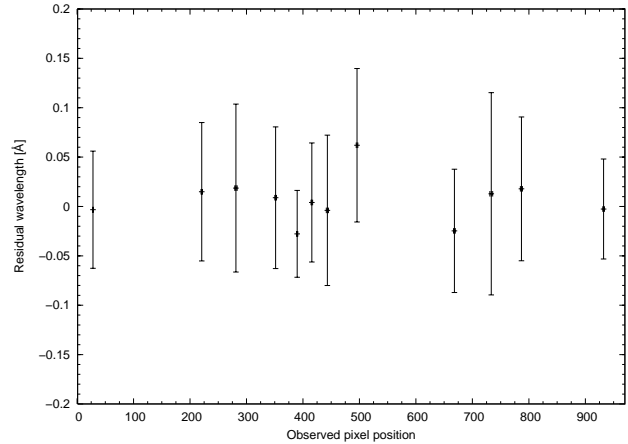}

\end{center}
\caption{{ Residuals from wavelength calibration} using the theoretical water vapour line positions measured from a HITRAN spectrum and the observed pixel positions of the corresponding features in the observed spectrum of our telluric standard star, \hbox{HD 75063}. The best fit was obtained using a cubic polynomial.}
\protect\label{fig:wavcalib}
\end{figure}

\subsection{Wavelength calibration}

At the time of observations, no arc lamp giving sufficient lines to perform a calibration in the 2.18 \micron\,-\,2.19 \micron wavelength range was available. We were thus unable to carry out a conventional calibration using the positions of known emission lines. Instead, we used a spectrum generated from a HITRAN line list \citep{rothman05hitran} to identify corresponding features in an observed spectrum of the telluric standard star, \hbox{HD 75063}. We used the simple emission line fitting routine {\sc emlt} in the Starlink {\sc figaro} package to fit Gaussians to the inverted spectra in order to identify the positions and width of the lines. The latter were used as uncertainties in the measurements. By fitting a cubic polynomial to the 12 identified telluric lines (Fig. \ref{fig:wavcalib}), we obtained a wavelength calibration which was subsequently used in our analysis. The \chisq~ of the fit using a cubic polynomial provided improvements by a factor of 31 and 1.8 when compared with a straight line and a quadratic fit respectively. { The 0.023 \AA\ rms residual scatter in the fit corresponds to \hbox{0.32 \kms}\ at the centroidal wavelength of 21484 \AA, or 0.06 of a resolution element.}.

\begin{figure*}
\begin{center}
\vspace{0.0cm}
\includegraphics[width=100mm,height=175mm,angle=270]{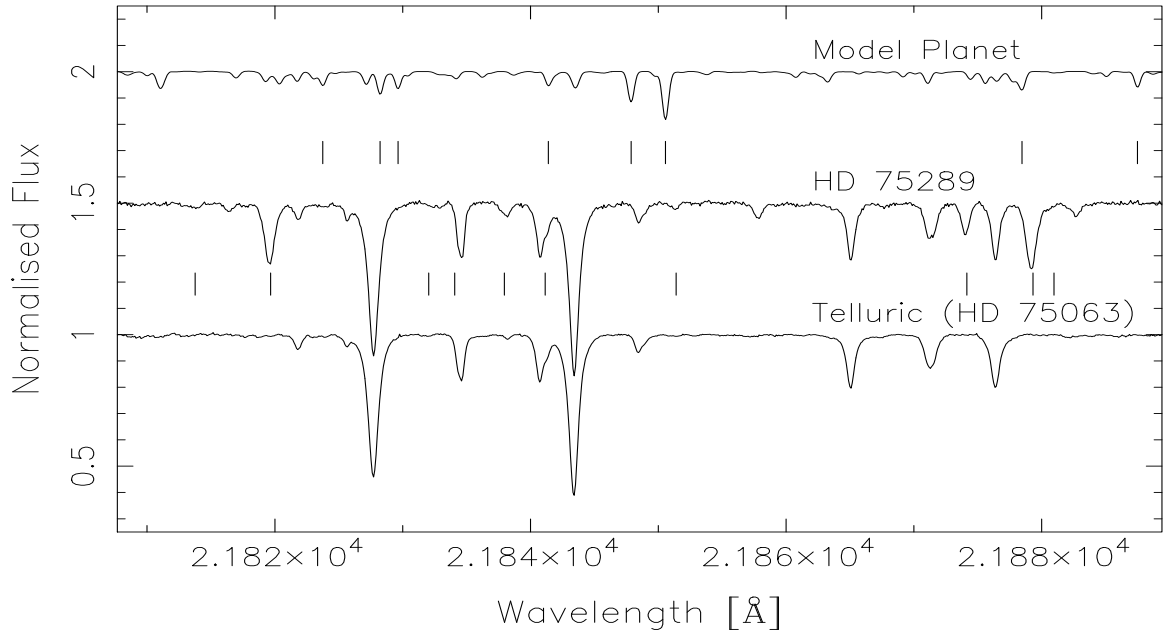}

\end{center}
\caption{{ Model and observed spectra in the 2.18\,-\,2.19 \micron~span of the observations at a resolution of R $\sim 56,800$. Top: Model planetary spectrum with tick marks below indicating the 8 strongest lines (see \S \ref{section:modelspectrum}). Middle: Observed HD 75289 spectrum plus undetected planetary spectrum and telluric features (first observation from January 16). Tick marks below indicate the theoretical positions of Fe and Si stellar lines as given by VALD (see \S \ref{section:modelspectrum}). There are additional absorption features not listed in VALD at $\lambda = 21816.4$~\AA, $21857.8$~\AA\ and $21882.8$~\AA. Bottom: HD 75063 A1\,III spectrum showing only telluric lines. The observed HD 75289 spectrum and model planet spectrum are offset by arbitrary units of 0.5 and 1.0 respectively for plotting purposes.}}
\protect\label{fig:planet_spectrum}
\end{figure*}

\subsection{Extracting the planetary signal}
\protect\label{section:extracting}
We first constructed a high S/N ratio master stellar spectrum template to accurately subtract the unshifted starlight from each observed spectrum. This has the additional benefit of removing (to first order) the telluric water vapour lines prevalent in this part of the spectrum. By monitoring the strengths of the telluric lines alone, we found, as expected, that they vary in strength throughout the night, generally being stronger when \hbox{HD 75289} was observed at high airmass. The telluric lines did not all behave in this way however, with some lines weaker than the mean at times while others were stronger. Using a single master frame comprising both stellar and telluric lines left residuals of up to 5-$\sigma$ { times the expected noise level} after subtraction of the starlight from each spectrum. 

{ Instead we made individual use of the bright star, \hbox{HD 75063}, which contains only telluric lines in the observed spectral range of our data. This star was observed at regular intervals on each night of observations. A mean telluric template spectrum was created for each night of observations by aligning and coadding all the \hbox{HD 75063} spectra observed throughout that night. The telluric spectrum was fitted to each \hbox{HD 75289} spectrum by taking derivatives of the spectra and using splines to calculate the scale factor at points across the spectra.} This process can account for lines which behave independently over the night (i.e. all telluric lines do not necessarily vary in strength by the same factor at any given time) and is described in \citet{cameron02upsand} (Appendix A). The procedure provides a spectrum which closely matches the telluric lines for each observed \hbox{HD 75289} spectrum and is used to divide out these features, leaving only stellar and planetary features.

A master stellar-plus-planet spectrum is then created for each telluric corrected \hbox{HD 75289} observation, but excluding the current telluric corrected \hbox{HD 75289} spectrum. Every stellar-plus-planet spectrum is then divided by an aligned and shifted version of the current master stellar-plus-planet spectrum in order to remove the stellar lines. Since a planetary signal is Doppler shifted according to phase relative to the stellar lines only the planetary signature should remain. { The master stellar-plus-planet spectrum which is subtracted from each spectrum in turn contains a blurred copy of the planetary signal. This will thus attenuate the planetary signal in the frame on which subtraction is being performed, to some degree. The effect is worst at $\phi = 0.25$ \& $\phi = 0.75$ where the planet moves very little in velocity space from spectrum to spectrum. The attenuation correction applied to the matched travelling Gaussian used to model the planetary motion when searching for a signal is described in detail in \S 6 and Appendices D2 \& D3 of \citet{cameron02upsand}.}

We found however that further time varying systematic residuals remained in the spectra and believe that these remaining residuals are due to the time varying ripple which we were unable to remove from the spectra at the extraction stage. Since we remove this effect using splines, we expect time varying differences in the fits which do not exactly match the observed changes in the ripple pattern due to finite signal to noise in the data. We therefore implemented a method which removes the residual signal which we treat as pattern noise. This method using principal component analysis is described by \citet{cameron02upsand} (Appendix B) and removes time varying patterns at fixed positions in the spectra. When the first few principal components are removed, this has little or no effect on any planetary signal since it changes position from spectrum to spectrum. We discuss the results using this algorithm further in section \S \ref{section:results}.

\subsection{Model spectrum}
\protect\label{section:modelspectrum}

The residual spectrum will contain a Doppler-shifted copy of the planetary spectrum, which at this stage is still buried in the noise. To { reduce the effective} noise, we model the planet's spectrum as the convolution of a Doppler-shifted mean line profile { (which we wish to determine)} and an absorption line pattern for an irradiated { \em model} atmosphere spectrum with parameters equal to those of \hbox{HD 75289b}. { Our model spectra were generated using the cloud-free ``rainout'' method described in BHA05 which improves the earlier models of \citet{barman01} (based on the "AMES-cond" models of \citet{allard01})} by iteratively reducing the elemental abundances involved in grain formation at a given layer and recomputing the chemical equilibrium at each new set of stratified elemental abundances. The resulting equilibrium chemistry and opacity sampling after rainout of species is fully self consistent, whereas earlier cond models simply excluded the grain opacities. Thus in the models used in this work, elements which are important in the atmospheres of cool stars such as Ti and V are significantly depleted leading to negligible concentrations of TiO and VO. Day-night temperature gradients were modelled under the assumption that concentric neighbouring rings of different temperature interact very little via radiative transfer processes. { For a fuller description of the model opacities and setup see \citet{allard01}, \citet{barman01} and BHA05. The temperatures on the dayside of the planet lead to an atmospheric chemistry dominated by H$_2$, He, H$_2$O and CO (BHA05). The dominant opacities in the 2.18\,-\,2.19 \micron\ wavelength span of our data however are due to H$_2$O. We derived the necessary list of line positions and depths for least squares deconvolution from a model spectrum before instrumental and stellar line broadening were added to the spectra and for the case when the planet is in conjunction with the star (i.e. at orbital phase 0.5).}

{ In Fig. \ref{fig:planet_spectrum} (top), we plot the model spectrum generated for \hbox{HD 75289b}. The spectrum has been convolved with a Gaussian to mimic the \hbox{R $\sim 56,800$} resolution of the observations and rotationally broadened to \hbox{1.482 \kms}\ under the assumption that the planet is tidally locked to the star \hbox{(i.e. \hbox{\vsini\ =} \mbox{$2 \pi R_p~{\rm sin}~i / P$})}. Fig. \ref{fig:planet_spectrum} also shows a typical \hbox{HD 75289} spectrum (middle) and \hbox{HD 75063} spectrum (bottom). The tick marks below then \hbox{HD 75289b} spectrum indicate the theoretical positions of 10 stellar lines (Fe and Si opacities) which are listed in the Vienna Atomic Line Database, VALD \citep{kupka99}. Additional opacities not included in VALD can also be seen (see Fig. \ref{fig:planet_spectrum} caption). The \hbox{HD 75289} spectra are thus clearly dominated by telluric lines.}

{ The mean absorption profile is recovered through use of a least squares deconvolution method first demonstrated by \citet{donati97zdi}. A list of absorption lines is derived from the {\em model} atmosphere (see above) and the depths, before any kind of broadening, are used to optimally weight each line in the {\em observed} spectrum. The deconvolution is effectively a sophisticated cross-correlation method which optimally coadds all the aligned absorption profiles while removing any side lobes due to line blending. The resulting profile is a mean absorption line with S/N ratio boosted, thereby improving the chance of detecting weak signals which are dominated by noise.} The current version of the code \citep{barnes98aper} { which propagates errors from the input spectra} has been used extensively for reflected light searches in the optical by \citet{cameron99tauboo,cameron02upsand} and \citet{leigh03a,leigh03b}. For \hbox{HD 75289b}, we expect { $\sim$98} planetary absorption features in the wavelength range of our data. { The weighted mean normalised depth \citep{barnes98aper} of these lines relative to the planetary ``continuum'' is 0.096. The optimal nature of the deconvolution means that all 98 lines in the wavelength range of our observations contribute, but only 8 (indicated by tick marks below the model planet spectrum in Fig. \ref{fig:planet_spectrum}) of the lines possess a normalised depth relative to the planetary continuum greater than the mean weighted value of 0.096. Of the 98 lines, 50 possess depths greater than 0.01 of the normalised planetary continuum. Attenuation by a further factor in the combined star+planet spectrum is one of the main parameters which we wish to determine.}

{ From photon statistics ($\sqrt{N_e}$ per spectrum, where $N_e$ = number of electrons), we expect an {\em mean} S/N limit in our spectra of \mbox{$\sim269$} for both nights of data. The mean S/N ratios, measured from flat regions in the normalised spectra, were \hbox{199$_{-166}^{+138}$ and} 268$_{-77}^{+70}$ for the nights of January 15 \& 16 respectively. The upper and lower limits represent the minimum and maximum recorded S/N ratio on each night. Slit alignment and tracking problems were the cause of low counts in a number of the January 15 spectra with a minimum S/N = 25 (readout noise = 40 e$^{-}$) . The gain in S/N of 3.4 for the deconvolved line profile yielded S/N ratios of 797 $_{-685}^{+352}$ over the two nights (i.e. minimum S/N $\sim$112 and maximum S/N = 1149).}

\section{Results}
\protect\label{section:results}

\subsection{Matched-filter analysis}
\protect\label{section:matchedfilt}

We model the time-dependent changes in Doppler shift and brightness, in the manner described in \S \ref{section:detect}. This matched-filter enables us to search for features in the timeseries spectra and is described in \citet{cameron02upsand} (appendix D). We asses the relative probabilities of the \chisq~fits to the data by varying $\epsilon(\lambda)$ and $K_{p}$ and plotting the improvement in \chisq, which is normalised to the best-fitting model. To calibrate any candidate detection, we construct a simulated planet signal of known $\epsilon_{0}(\lambda)$ that is added to the extracted spectra prior to { removal of stellar and telluric features and before least squares deconvolution is carried out}. By ensuring the fake planet is recovered correctly by our procedures, we can be confident of a calibrated detection in the presence of a genuine planet signal. 

The significance of the result is assessed using bootstrap statistical procedures based on the random re-ordering of the data in a way that scrambles phases while preserving the effects of any correlated systematic errors \citep{cameron02upsand}. The order of the observations is randomised in a set of 3000 trials which will scramble any true planetary signal while enabling the data to retain the ability to produce spurious detections through the chance alignment of systematic errors. The least squares estimate of log$_{10}\,\epsilon(\lambda)$ and associated \chisq~as a function of $K_p$ enable us to plot 68.4, 95.4, 99.0 and 99.9 per cent bootstrap limits on the strength of the planetary signal.

\begin{figure*}
 \begin{center}
   \begin{tabular}{cc}
      \vspace{10mm} \\
      \includegraphics[width=70mm,bbllx=113,bblly=114,bburx=397,bbury=397,angle=0]{barnes_hd75289_fig6.eps} \hspace {5mm} &
      \hspace {5mm}
      \includegraphics[width=70mm,bbllx=113,bblly=114,bburx=397,bbury=397,angle=0]{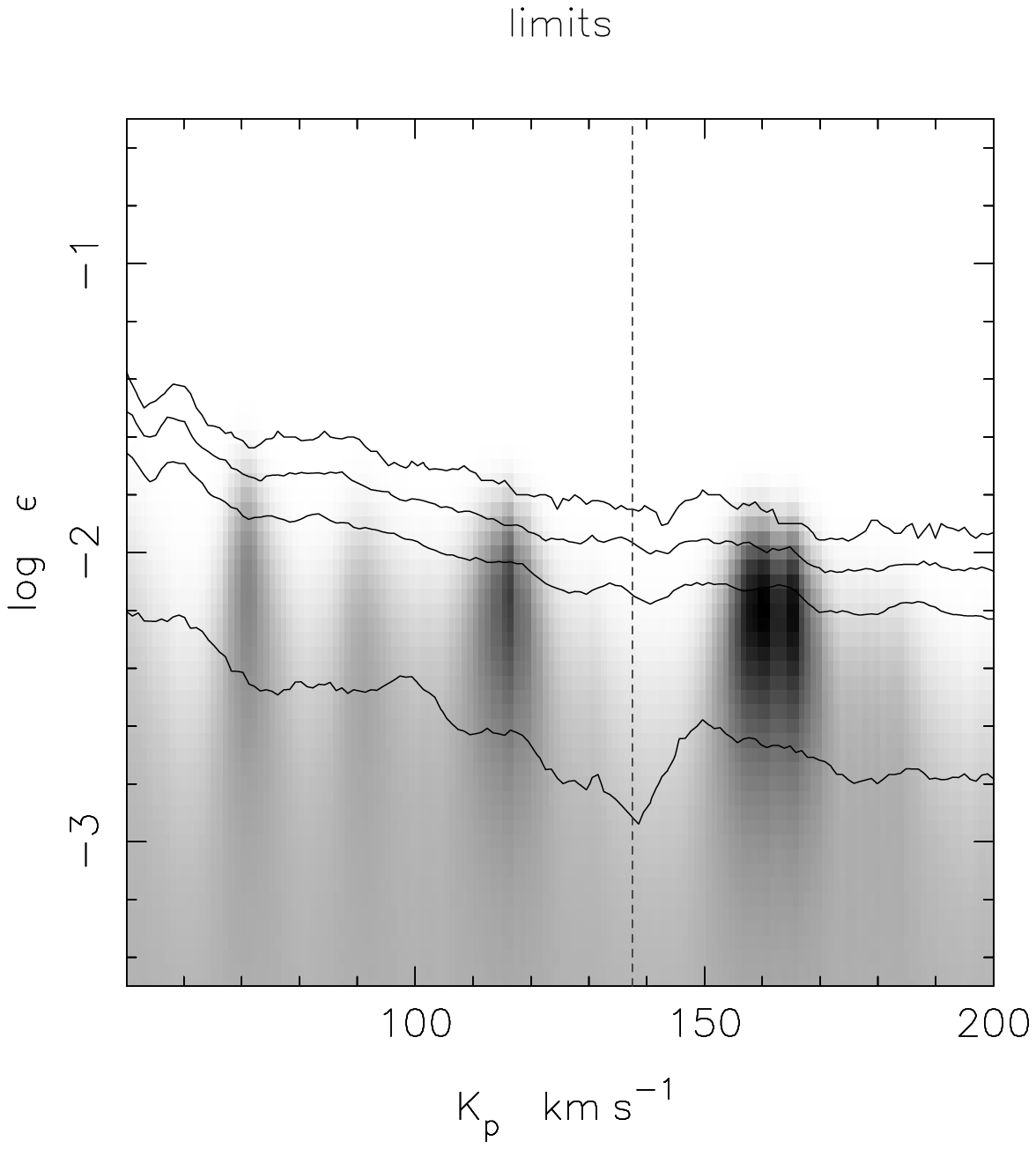} \\

      \vspace{20mm} \\
      \includegraphics[width=70mm,bbllx=113,bblly=114,bburx=397,bbury=397,angle=0]{barnes_hd75289_fig8.eps} \hspace {5mm} &
      \hspace {5mm}
      \includegraphics[width=70mm,bbllx=113,bblly=114,bburx=397,bbury=397,angle=0]{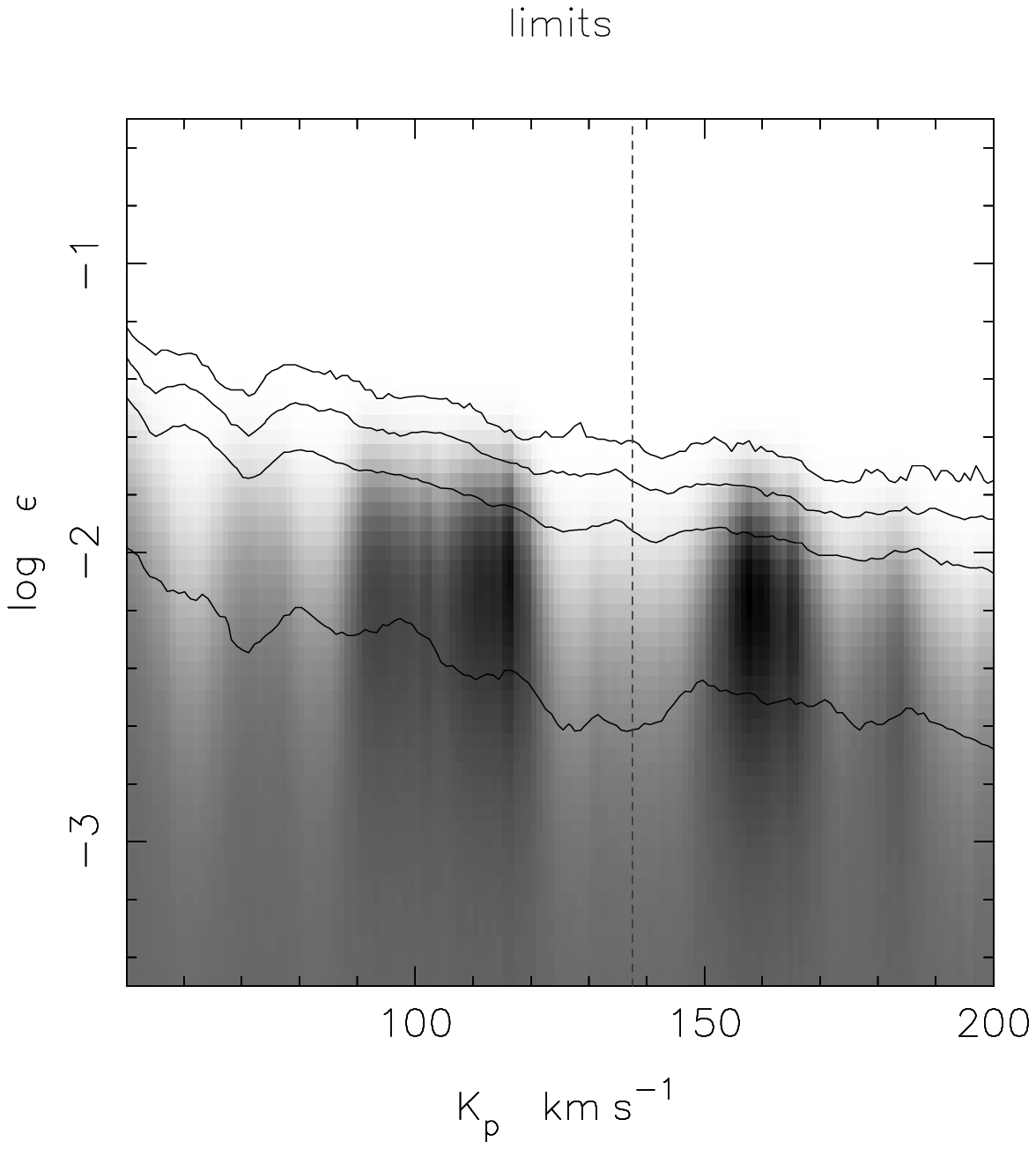} \\
      \vspace{10mm} \\

   \end{tabular}
 \end{center}

\caption{Left: Phased timeseries of the deconvolved residual spectra of \hbox{HD 75289b}. { The dashed sinusoidal curve represents the motion of a planetary signal based on the {\em most probable} velocity amplitude, \mbox{\^{\em K}$_{p} = 137.55$}~\kms}, estimated from empirically determined parameters (\S \ref{section:searching}) { and does not represent a detection with this amplitude}. Right: Relative probability \chisq\ map of planet-star flux ratio $log_{10}\,\epsilon(\lambda)$ vs $K_p$. Results are shown after removal of the first two (top) and first five (bottom) principal components from the timeseries spectra. The greyscale in the probability maps represents the probability relative to the best fitting model in the range white for 0 to black \hbox{for 1}. Plotted are the 68.4, 95.4, 99 and 99.9 per cent (bottom to top) confidence levels { for a planetary signal detection.} { The dashed vertical line represents the most probable velocity amplitude.} The dark feature at $K_p$ = 157.8 \kms~and \hbox{$log_{10}(\epsilon)$ = -2.2} is detected with { $< 95.4$ per cent} confidence. { Note how removal of five principal components reduces the significance of this feature such that it becomes approximately equal in significance to the \hbox{$K_p$ = 116.5 \kms}\ candidate (bottom right). These features are close to the noise level and are probably due to incomplete removal of the fixed pattern seen in the timeseries spectra.}}
\protect\label{fig:data_planet}
\end{figure*}

\subsection{Searching for a planetary signal}
\protect\label{section:searching}

Fig. \ref{fig:data_planet} (left) shows the phased timeseries deconvolved spectra and the resulting \chisq~map of log$_{10}\,\epsilon(\lambda)$ vs $K_p$. The effective wavelength after performing deconvolution is \hbox{$\lambda = 21848$ \AA}. Fig. \ref{fig:data_planet} (right ) shows the relative probability \chisq\ map of the planet-star flux ratio $log_{10}\,\epsilon(\lambda)$ vs $K_p$. The { darkest regions in the map represent the greatest improvement in \chisq\ when fitting the model described in \S \ref{section:matchedfilt}. We can obtain an estimate of the upper limit \hbox{(i.e. when $i = 90$\degs)} of the velocity amplitude of the planet.  With \mbox{$a = 0.0482 \pm 0.0028$ AU} and \mbox{$P = 3.509267 \pm 0.000064$ d}, we find \mbox{$v_p = max(K_p) = 149.43 \pm 8.68$}~\kms. With the orbital inclination of \hbox{$i = 67$}\degs\ reported in \S \ref{section:hd75289}, we estimate a most probable velocity amplitude of \mbox{\^{\em K}$_{p} = 137.55 \pm 7.99$}~\kms. In \hbox{Fig. \ref{fig:data_planet}} we plot the corresponding motion and recovered position for \^{\em K}$_{p}$ (dashed curves and vertical lines respectively) as a visual guide to the reader.}

It is possible to see the non-Gaussian systematics in the phase plot in Fig. \ref{fig:data_planet} { (top left)}. We believe that these features are the result of our inability to fully remove the varying telluric signatures { and uncharacterised changes in the observed ripple function described in \S \ref{section:data_reduction}} . To remove the most significant trends in the { timeseries data} at fixed positions in wavelength, we used principal component analysis { (\citet{cameron02upsand}, Appendix B). The eigenvalues of the first two principal components indicated a strong contribution while a less strong contribution from the next five components was then followed by a linear decrease in significance.} We thus investigated removal of the first two, and the first five components, with the resulting input timeseries for the latter case being significantly cleaner of systematic noise features. { When we compare Fig. \ref{fig:data_planet} (bottom right) with Fig. \ref{fig:data_planet} (top right), we see how the \hbox{$K_p = 157.8$ \kms}\ feature is reduced in significance, leaving several features of comparable strength and low statistical significance. The \hbox{$K_p = 157.8$ \kms}\ feature only marginally lies in the range of possible values since the uncertainty on our $v_p = max(K_p)$ measurement yields an upper (1-$\sigma$) limit of \hbox{158.11 \kms}. A planet with this velocity amplitude would imply $i \sim 86.5$\degs, indicating a planet likely to show transits, an observation not reported in the literature. The \hbox{$K_p = 116.5~$\kms}\ feature is of no greater significance and implies an orbital inclination of \mbox{$i = 47.5$}\degs\ respectively. This is unlikely to give a detection given the relatively low maximum flux we would receive under the reasonable assumption \citep{harrington06} that heat is not effectively redistributed and the planet is hottest on its dayside. All candidate features however can confidently be rejected on the basis of the observed flux ratios for a number of systems (\S \ref{section:intro}) since the mean \hbox{log$_{10}\,\epsilon(\lambda)$ = -2.1} implies a planet significantly hotter than even \hbox{HD 189733} (see Fig. 8 of \citet{fortney06hd149026b}).

Under the assumption that all candidate features are spurious, the overall level of noise in the data enables us to rule out the presence of a planet at \^{\em K}$_p$ with log$_{10}\,\epsilon_{0}(\lambda) > -2.1$ at the 2-$\sigma$ level (Fig. \ref{fig:data_planet}, top right).}

\begin{figure*}
 \begin{center}
   \begin{tabular}{cc}
      \vspace{10mm} \\
      \includegraphics[width=70mm,bbllx=113,bblly=114,bburx=397,bbury=397,angle=0]{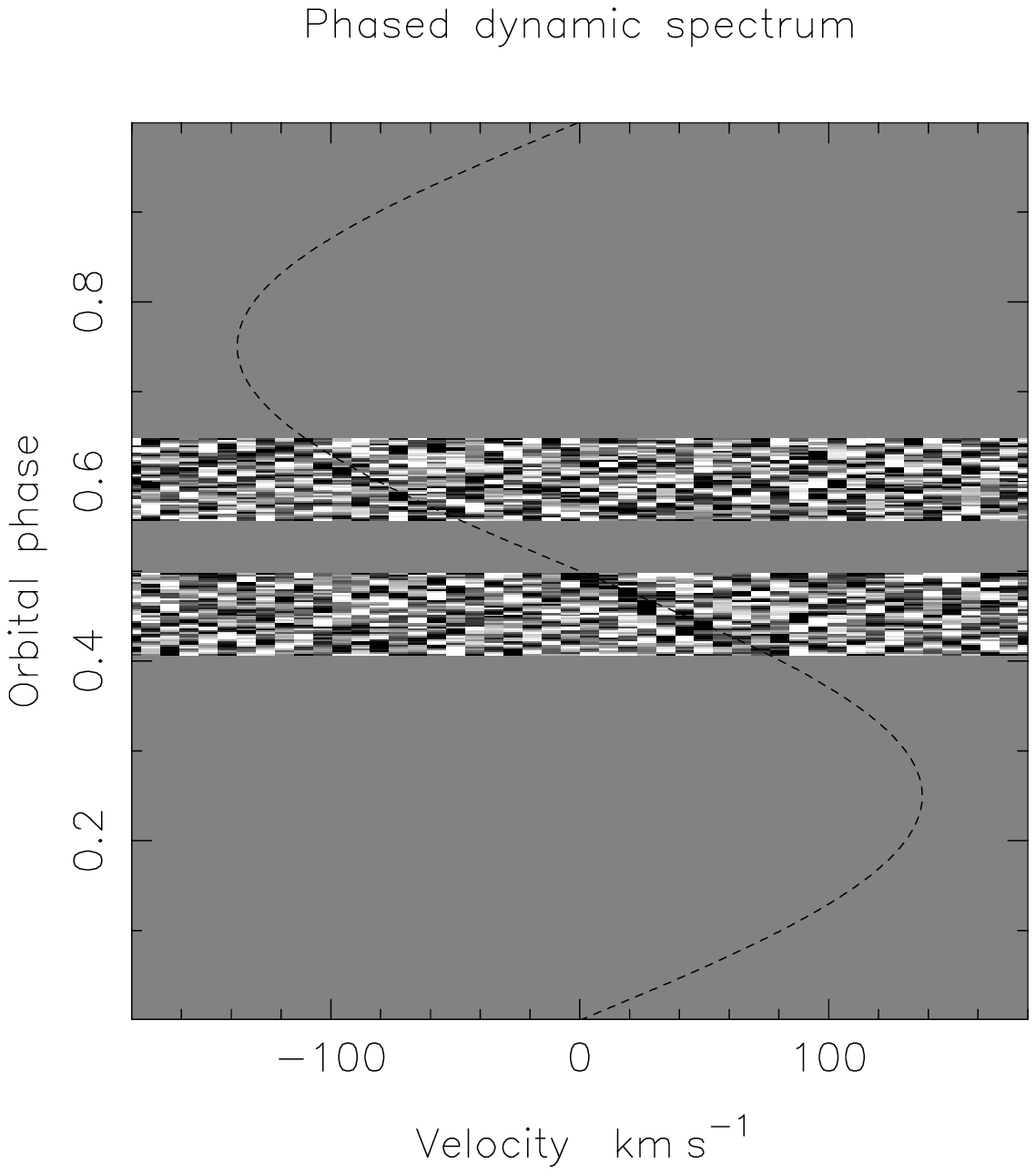} \hspace {5mm} &
      \hspace {5mm}
      \includegraphics[width=70mm,bbllx=113,bblly=114,bburx=397,bbury=397,angle=0]{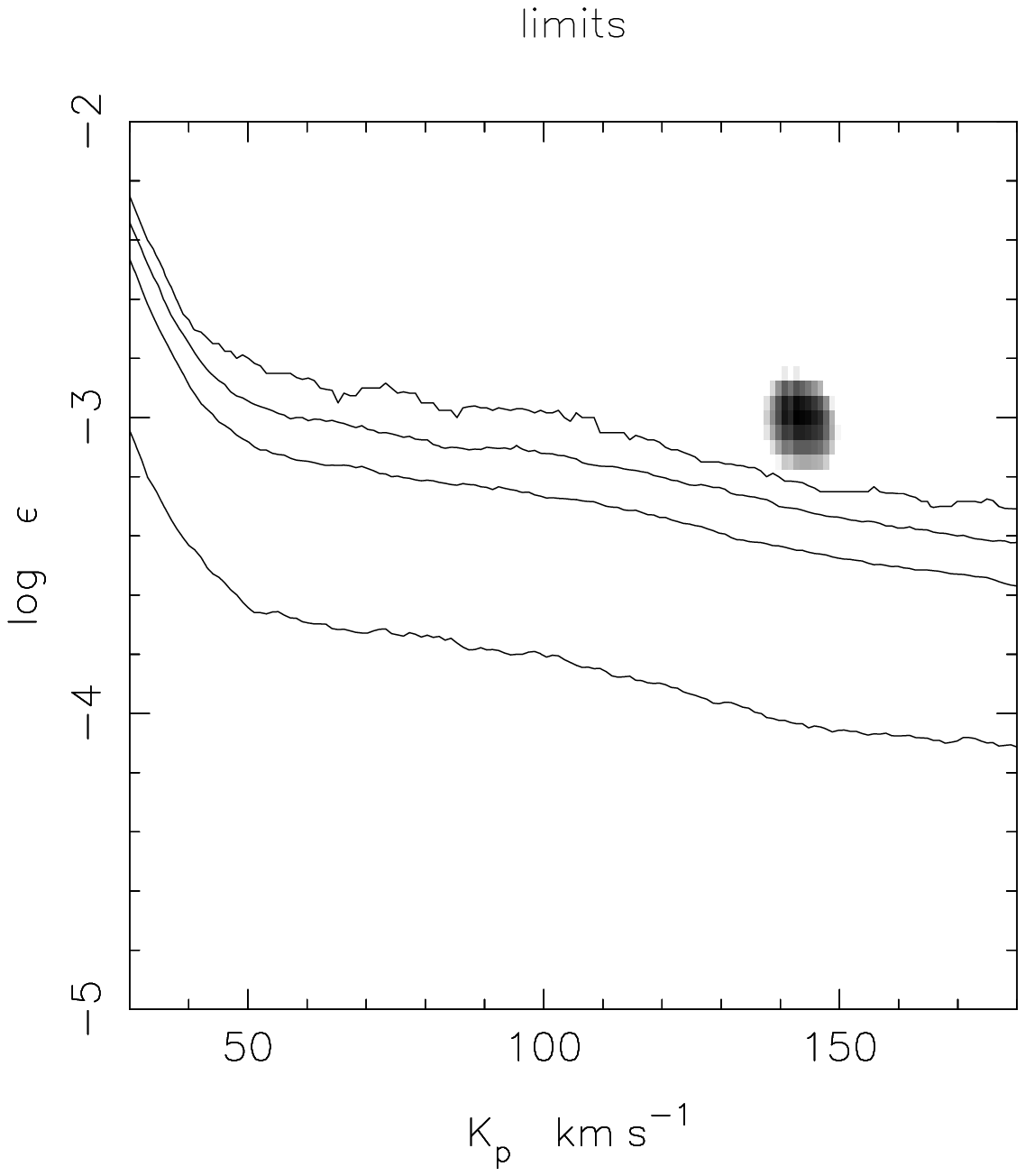} \\

      \vspace{20mm} \\
      \includegraphics[width=70mm,bbllx=113,bblly=114,bburx=397,bbury=397,angle=0]{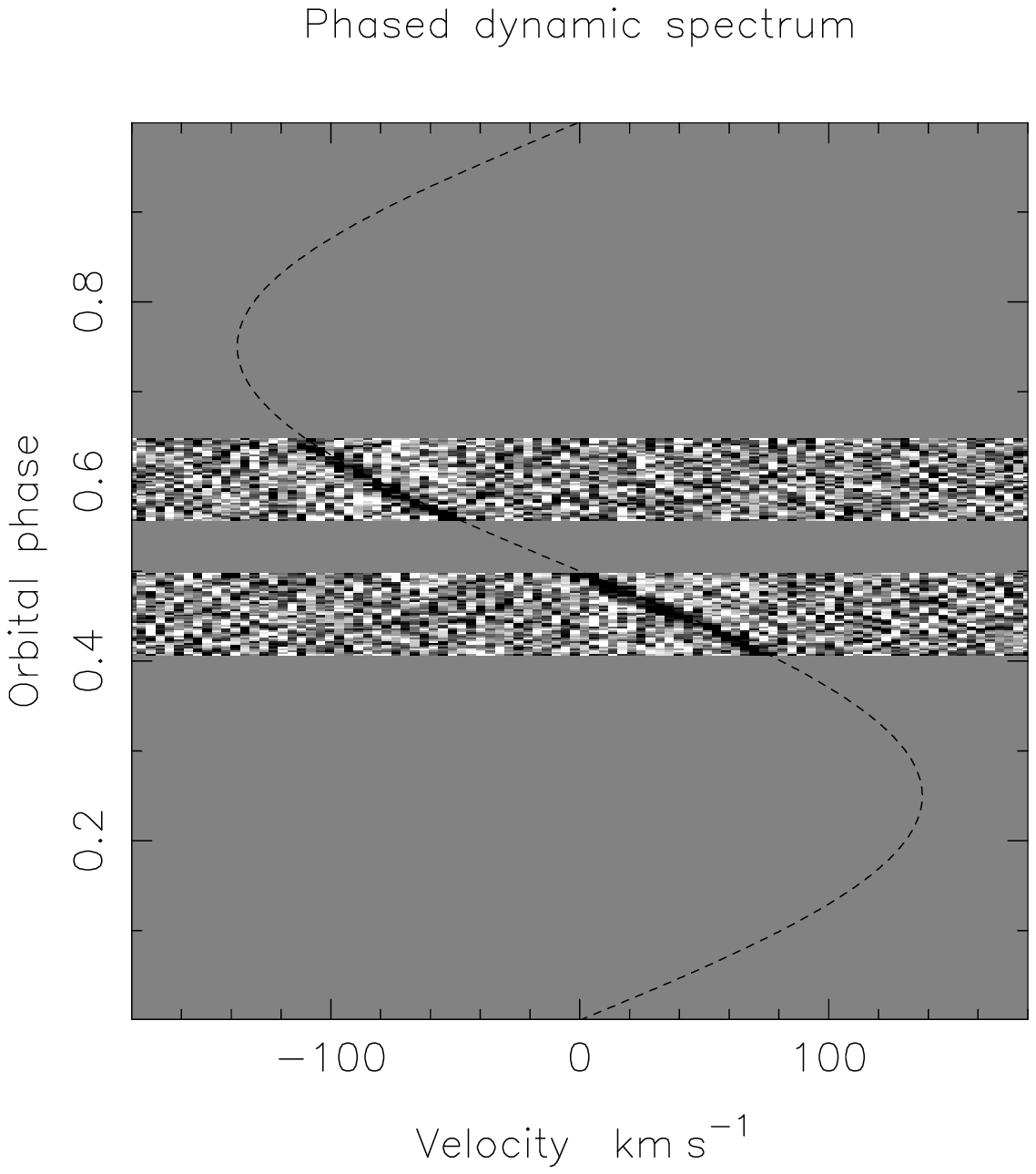} \hspace {5mm} &
      \hspace {5mm}
      \includegraphics[width=70mm,bbllx=113,bblly=114,bburx=397,bbury=397,angle=0]{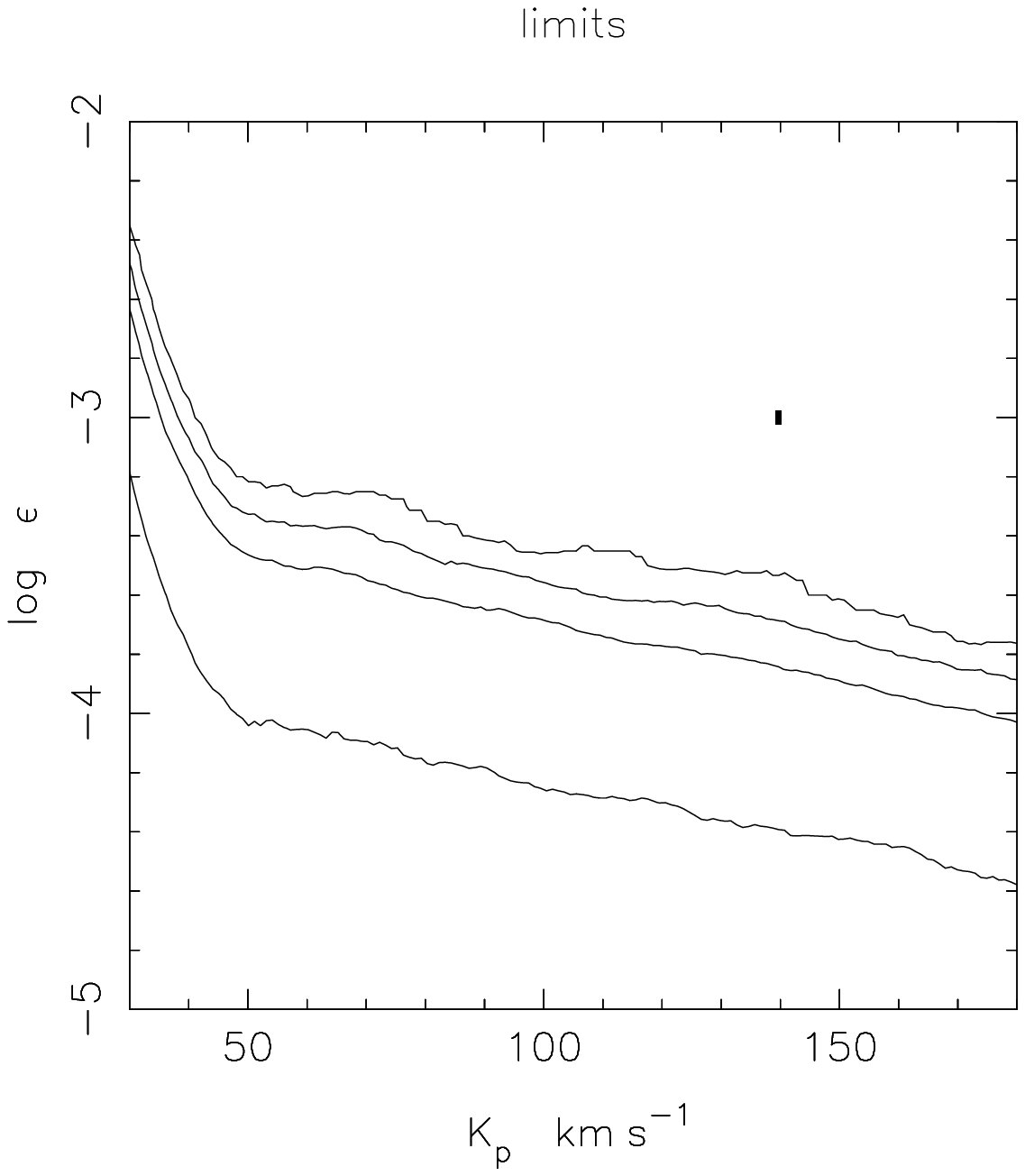} \\
      \vspace{10mm} \\

   \end{tabular}
 \end{center}

\caption{As for Fig. \ref{fig:data_planet} but using simulated data (50 spectra taken at two epochs) with a mean S/N ratio of 300. The dashed sinusoidal curve represents the {\em known} $K_p$ of the simulated planet. The spectra were simulated to mimic the multi-order capability of existing cross-dispersed spectrographs, covering 7 orders and spanning the region of 1.90 \micron~to 2.45 \micron. Resolutions of R $\sim$20000 (top) and R $\sim$40000 (bottom)  were used. { A detection at a level of $\sim$ 3 times the  99.9 per cent { confidence} level is achieved, indicating a limiting detection (with the same confidence) for spectra obtained with a mean \hbox{S/N $\sim$ 170}}. The greyscale in the \chisq\ plots runs from 0.99 (white) to 1.00 (black).}
\protect\label{fig:sim_planet}
\end{figure*}

\section{Simulating a planetary signal}
\protect\label{section:simplanet}

In order to assess the feasibility of detecting a planetary signature in the NIR, we have carried out simulations using fake data based on the known physical parameters of \hbox{HD 75289b}. By considering the wavelength coverage possible with current NIR multiorder cross-dispersed \'{e}chelle technology (i.e. IRCS/Subaru \citep{kobayashi00ircs} and NIRSPEC/Keck 2 \citep{mclean98nirspec}), we estimate that approximately 50 spectra with R $\sim 20000$, a mean S/N ratio of 300 and encompassing a wavelength range of 1.90 \micron~to 2.45 \micron~could be obtained per \hbox{8 hrs} of observations \citep{kobayashi00ircs}. With a typical spectral coverage of 70\% of this range our planetary model atmosphere for HD 75289 yields $\sim2300$ useful lines with depths between 0.05 and 1.0 of the normalised continuum.

We generated model spectra for a G0V star using \hbox{ATLAS9} models \citep{kurucz93cdrom} and a HITRAN model of the telluric lines \citep{rothman05hitran}. These spectra were combined with our model planetary spectrum assuming an orbital inclination of \hbox{$i = 67$\degs}, a relative maximum planet-star flux ratio of \hbox{log$_{10}\,\epsilon_0 = -3$} and a velocity amplitude of \hbox{$K_p = v_p sin i$ $= 137.6$~\kms}. Gaussian random noise was added to the spectra to simulate a S/N ratio of 300. We assumed observations were taken over two nights with a ten night phase gap. This strategy optimises observations at phases close to $\phi = 0.5$, while minimising the timebase.

We attempted to recover of the planetary signal in the same way as described in previous sections. The phased timeseries spectra are presented in Fig. \ref{fig:sim_planet} (top left), and show a detection well clear of the 99 per cent confidence level (Fig. \ref{fig:sim_planet}, top right). Since the semi-major axis of the orbit and the period are both known, we can use our estimate of $v_p sin i$ to determine the inclination. Our matched-filter analysis gives $K_p = 142.7$~\kms\ leading to an inclination determination of 72\degs. Hence we overestimate the inclination by 5\degs. The detection limit at the simulated velocity of the planet is log$_{10}\,\epsilon_0 = -3.2$. Equivalently, at the projected velocity of the planet, we expect to recover the signal at the limiting confidence for data with S/N $\geq$ 200. 

We carried out a further test to estimate the feasibility of recovering the planetary signal in the case where the line strengths are incorrect. We assumed that the positions of all lines in our input model planetary spectrum could be incorrect by some fraction. The depth of each line, but not the position, was modified by a Gaussian random fraction of the original strength. We then attempted to recover the planetary signal with the modified line list (used for deconvolution) and found that for data with S/N = 300, we still recover the planet with a 99.9 per cent confidence limit for strengths which have been modified by on average 50 per cent. This simulation essentially mimics the effects of non-optimal extraction of the lines using our deconvolution code. Our ability to recover the planet is more sensitive to incorrect positioning of lines however. We find that a clear 99.9 per cent confidence detection is achieved in the limiting case of randomly re-positioning 15 per cent of the lines before deconvolution. The main reason these mismatches, which we discuss in the next section, is likely the uncertainties in opacities, opacity oscillator strengths and atmospheric physics of the planet. We note that in reality, removal of the telluric lines at the shorter wavelengths of the above range \hbox{(i.e. 1.90 \micron}~to 2.08 \micron) does not affect our ability to make a clear detection.

We also simulated recovery of the planet for a spectral resolution of R $\sim 40000$. The main advantage of high resolution data would be the increase in depth of the absorption lines relative to the continuum. The results are shown in Fig. \ref{fig:sim_planet} (bottom) and show the clear gain attained by doubling the resolution. Our limiting 99 per cent confidence threshold in this instance would enable us to detect a planetary signal at a level of log$_{10}\,\epsilon_0 = -3.5$. The planet is recovered with $K_p = 139.7$~\kms, leading to an inclination determination of 69.2\degs. Since a narrower slit will be used to make observations at higher resolution, we may not observe this degree of improvement in reality. Our \hbox{limiting 99.9 per cent} confidence indicates we would still detect the planet with \hbox{S/N = 95} or approximately 1/4 the number of photons required in the R $\sim 20000$ limiting case. We discuss the prospect of obtaining spectra of this nature in section \S \ref{section:discussion}

\section{Discussion}
\protect\label{section:discussion}

The methods employed in this paper provide a very encouraging indication that a planetary signal can potentially be extracted from high resolution timeseries spectra. The fact that we do not detect a signal can be attributed largely to systematic noise introduced from difficulties in consistently normalising each spectrum in our timeseries. This arises largely because of the systematic and varying large scale flat fielding errors associated with the detector employed in the observations as discussed in \S \ref{section:data_reduction}.

We have shown that the feature detected in our spectra is unlikely to be a real planetary feature since its strength is much reduced when additional principal components, which describe (systematic) trends in the data, are removed. The eigenvector describing the contribution of principal components shows a very strong contribution from the first 2 components with an exponential decay which levels off after the 7th principal component. At this level, the detected feature becomes consistent with the shot noise in the spectra. The performance of NIR detectors is clearly of great importance if we wish to detect small signals dominated by a nearby bright object. More modern detectors such as the 1024$^2$ \hbox{Aladdin III} array used in conjunction with NIRSPEC at Keck\ 2\ \citep{mclean98nirspec} do not show the characteristics \citep{mclean00} of the older \hbox{Aladdin II} array used to secure the data in this work. Future detectors promise even greater stability with \cite{bezawada06} finding 0.05 per cent flatfield stability over a period of one day for a 2048$^2$ HAWAII-1RG engineering array.

Current observational evidence for a planetary signal at \hbox{2.2 \micron}~therefore remains inconclusive. Initial attempts to search for extrasolar planetary spectra (by the method of difference analysis where the spectrum of the host star seen during eclipse of the planet is subtracted from the spectrum of the combined star and planet, seen out of eclipse) revealed null results. The analysis by \citet{richardson03} failed to detect the predicted continuum `bump' at  2.1\,-\,2.2 \micron\ at a level of log$_{10}(F_p/F_*)= -3.52$. \citet{snellen05} did not detect the eclipse shape from 2.3\micron~secondary eclipse observations of \hbox{HD 209458b}, but nevertheless the mean drop in flux (based on two measurements) of 0.115 $\pm$ 0.139 per cent indicates a planet-star flux ratio of 0.0010 $\pm$ 0.0009. While this confirms model predictions for a planet-star flux ratio, \citet{deming06ir} note that the size of the uncertainty does not exclude the blackbody result.

While advances have been made in detecting spectral features in the mid-infrared 7.5\,-\,13.2\micron~interval \citep{richardson07spectrum,grillmair07spectrum} using Spitzer space telescope observations, models would suggest that the higher contrast ratio in the NIR will necessitate the kind of signal enhancement techniques used in this work. It is clear that there is still much to learn about the atmospheres of CEGPs. Both \citet{richardson07spectrum} and \citet{grillmair07spectrum} have found that there is little evidence for the H$_2$O opacities present in model spectra. \citet{richardson07spectrum} however find evidence for silicate clouds in the spectrum of \hbox{HD 209458b}.

The simulations we have carried out in this work make the assumption that the opacities in current models are correct, even if their strengths are not. If many species are omitted or are simply not present in reality, this could have significant impact on the detection method, either for better or worse. As with the \citet{richardson03} search for the 2.1\,-\,2.2 \micron\ bump, our models contain unreliable OH opacities and oscillator strengths and missing CH$_4$ opacities which are seen in T dwarf spectra \citep{burgasser06}. This is clearly important when using methods which aim to detect the `continuum' shape, while we found (\S \ref{section:simplanet}) that incorrect positioning of greater than 15 per cent of lines would affect our ability to make a clear detection. Conversely, competing models could be used to obtain an optimal extraction of the absorption signature of a planetary spectrum. Switching off model opacities for species which are known to have unreliable positions may be necessary to enable recovery of the planetary signal. In the case of a clear detection, it should then also be possible to distinguish between differing phase function models by monitoring the improvement in \chisq~in our matched filter analysis.

If the albedo spectrum is largely independent of wavelength in the regions where absorbing species do not play a { significant} role, including 2.2 \micron\ (see Fig. 8 of \citet{sudarsky00}), we would expect a very small reflected light component in the NIR, given the upper limits found in the optical. Our data contain only { 10} lines arising from the spectrum of \hbox{HD 75289} itself, with a mean depth { 2.7 times} the mean depth of the { 98} lines we expect to see in the planetary atmosphere. {Nevertheless, if the 10} lines are seen as a component reflected from possible clouds in the atmosphere of \hbox{HD 75289b}, we expect little gain in the signal if the upper albedo limit of 0.12 \citep{leigh03b} is taken into account. The maximum reflected light signal in this case is only log$_{10}(F_p/F_*)= -4.4$, over an order of magnitude less than the { NIR} log$_{10}(F_p/F_*)\sim -3.0$ { estimate} adopted throughout this work. 

\subsection{Summary}
\protect\label{section:summary}

We have presented an analysis and feasibility study for the detection of the NIR spectroscopic signature of the close orbiting extrasolar giant planet, \hbox{HD 75289b}. We do not detect the planet with the present Gemini/Phoenix data but through simulations based upon models and observational constraints we expect that current cross-dispersed spectrographs, such as IRCS/Subaru \citep{kobayashi00ircs} and NIRSPEC/Keck\ 2 \citep{mclean98nirspec}, operating at resolutions in the region of R $\sim 20000$ are sufficient to obtain a detection with 99.9 per cent confidence. Facilities which could offer higher resolution and broad wavelength coverage will be an enormous benefit. With a large spectral capability, it will be possible to determine the NIR spectral energy distribution in J, H \& K regions, complementing observations currently being made at longer wavelengths.

\section{Acknowledgments}
{ This paper is based on observations obtained with the Phoenix infrared spectrograph, developed and operated by the National Optical Astronomy Observatory. We would like to thank the anonymous referee for useful comments.} JRB was supported by a PPARC funded research grant during the course of this work.


\end{document}